\documentclass{article}
\usepackage{spconf,amsmath,graphicx}
\usepackage[hidelinks]{hyperref}
\usepackage{booktabs}
\usepackage{tabularx}
\usepackage{array}
\usepackage{multirow}
\usepackage{makecell}
\newcolumntype{Y}{>{\raggedright\arraybackslash}X}

\title{ADEPT: AN ENTROPY-DRIVEN DUAL-STRATEGY AGENT FOR INTERACTIVE VIDEO RETRIEVAL}
\name{Ke Chen$^{1,*}$, Shengyuan Han$^{2,*}$, Yongfeng Huang$^{3}$, Jundong Liu$^{4,\dagger}$, Jingwei Xiong$^{5,\dagger}$, Liang Xu$^{6,\dagger}$, Yujin Zhu$^{6}$}

\address{
$^{1}$Institute of Science Tokyo, Japan \\
$^{2}$Nanjing University of Posts and Telecommunications, Jiangsu, China \\
$^{3}$The Chinese University of Hong Kong, Hong Kong SAR \\
$^{4}$City University of Hong Kong, Hong Kong, China \\
$^{5}$University of California, Davis, CA, USA \\
$^{6}$Artificial Intelligence Innovation Center, Yangtze Delta Region Institute of Tsinghua University, China \\
{\small $^{*}$ Equal contribution (co-first authors). $^{\dagger}$ Corresponding authors.}
}

\begin{document}
%
\maketitle
\begin{abstract}
This research aims to solve the challenge of video retrieval from massive datasets, caused by ambiguous user queries. Prevailing single-round retrieval paradigms face a performance bottleneck, as they lack effective feedback mechanisms to handle complex search intentions. The root cause is the ``Intent-Query Gap'', where users' intent cannot be captured by a simple text query. To solve this, we propose the \textit{ADEPT} framework: a training-free agent that pioneers an entropy-driven decision engine to efficiently guide dialogue by dynamically selecting between ASK and REFINE strategies. Experiments on two challenging datasets demonstrate that ADEPT significantly outperforms all non-interactive, heuristic, and Video-LLM baselines. The core contribution of this work is an efficient and interpretable entropy-driven interactive strategy that sets a new performance benchmark for the field of interactive video retrieval.
\end{abstract}
\section{Introduction}
\label{sec:intro}

The exponential growth of short-form video creates a significant ``Intent-Query Gap'' where users' vague episodic memories clash with the ambiguity of textual queries. While interactive retrieval shifts the paradigm from static matching (e.g., CLIP4Clip~\cite{luo2021clip4clip}) to multi-turn dialogue, it exposes a critical \textbf{``Strategy Gap''}: the lack of an optimal policy for \textit{when} and \textit{what} to ask. Existing solutions relying on complex agents or rigid heuristics~\cite{shang2024traveler, han2024merlin, fan2024videoagent, maeoki2020interactive, liang2023simple} often lack transparency and efficiency.

To bridge this gap, we propose \textbf{ADEPT}, a lightweight, training-free framework powered by an information-theoretic engine. By dynamically diagnosing retrieval uncertainty via \textit{inter-cluster} ($H_{inter}$) and \textit{intra-cluster} ($H_{intra}$) entropy, ADEPT adaptively arbitrates between multi-granularity \textit{Ask} and \textit{Refine} strategies. This offers an interpretable, zero-shot solution to the intent disambiguation problem. Experiments on MAFW~\cite{liu2022mafw} and MER2024~\cite{lian2024mer} confirm that ADEPT significantly outperforms non-interactive and general-purpose Video-LLM baselines.

The core contributions of this research are as follows:
\begin{itemize}
    \item We propose the ADEPT framework, featuring a core entropy-driven engine that dynamically adapts its dialogue strategy in a zero-shot manner.
    \item We design an efficient and interpretable dual-strategy protocol (ASK/REFINE) that resolves multi-level query ambiguities through multi-granularity questioning.
\end{itemize}

\section{Related Works}
\label{sec:format}

Interactive video retrieval bridges the ``Intent-Query Gap'' through multi-turn dialogue. This shifts the research paradigm from static semantic matching to defining an optimal dialogue policy for the ``Strategy Gap''. Current approaches are mainly categorized into non-interactive retrieval, heuristic-based interaction, and agent-based systems driven by Video LLMs.

Current approaches can be broadly categorized as follows:
\begin{itemize}
    \setlength{\itemsep}{0pt}
    \setlength{\topsep}{0pt}
    \setlength{\parsep}{0pt}
    \item \textbf{Non-interactive Retrieval: }In single-round text-video matching, foundational work like CLIP4Clip~\cite{luo2021clip4clip} adapted CLIP for video, while subsequent works improved performance using techniques such as multi-grained contrastive learning and large language models~\cite{ma2022x, cheng2024emotion, thawakar2024composed}.
    \item \textbf{Heuristic-based Interactive Retrieval: }This approach has evolved from dataset-specific supervised question generators~\cite{maeoki2020interactive} to recent approaches using pre-trained VideoQA models with fixed rules to simulate interaction~\cite{liang2023simple}.
    \item \textbf{Video LLMs: }The research frontier has shifted to LLM-based cognitive agents, with approaches like TraveLER's~\cite{shang2024traveler} multi-agent framework, query reformulation, and embedding space navigation~\cite{han2024merlin, fan2024videoagent}. These are often benchmarked against powerful zero-shot models like LLaVA-NeXT-Video~\cite{li2024llava}.
\end{itemize}

\section{Methodology}
\label{sec:pagestyle}

We propose ADEPT, a training-free framework that interprets user intent through entropy-driven interactions.

\subsection{Overview of the Framework}
\label{ssec:subhead}
Given a database $\mathcal{D}$ and query $D_0$, the system iteratively refines results over $T$ rounds. We map inputs to a shared semantic space via an encoder $f_{\text{enc}}(\cdot)$. At round $t$, the query $D_t$ retrieves the candidate set $C_t = \text{Top-}K_{v \in \mathcal{D}} (\text{SIM}(\mathbf{q}_t, f_{\text{enc}}(v)))$. 

\begin{figure*}[t]
  \centering
  \includegraphics[width=0.8\textwidth]{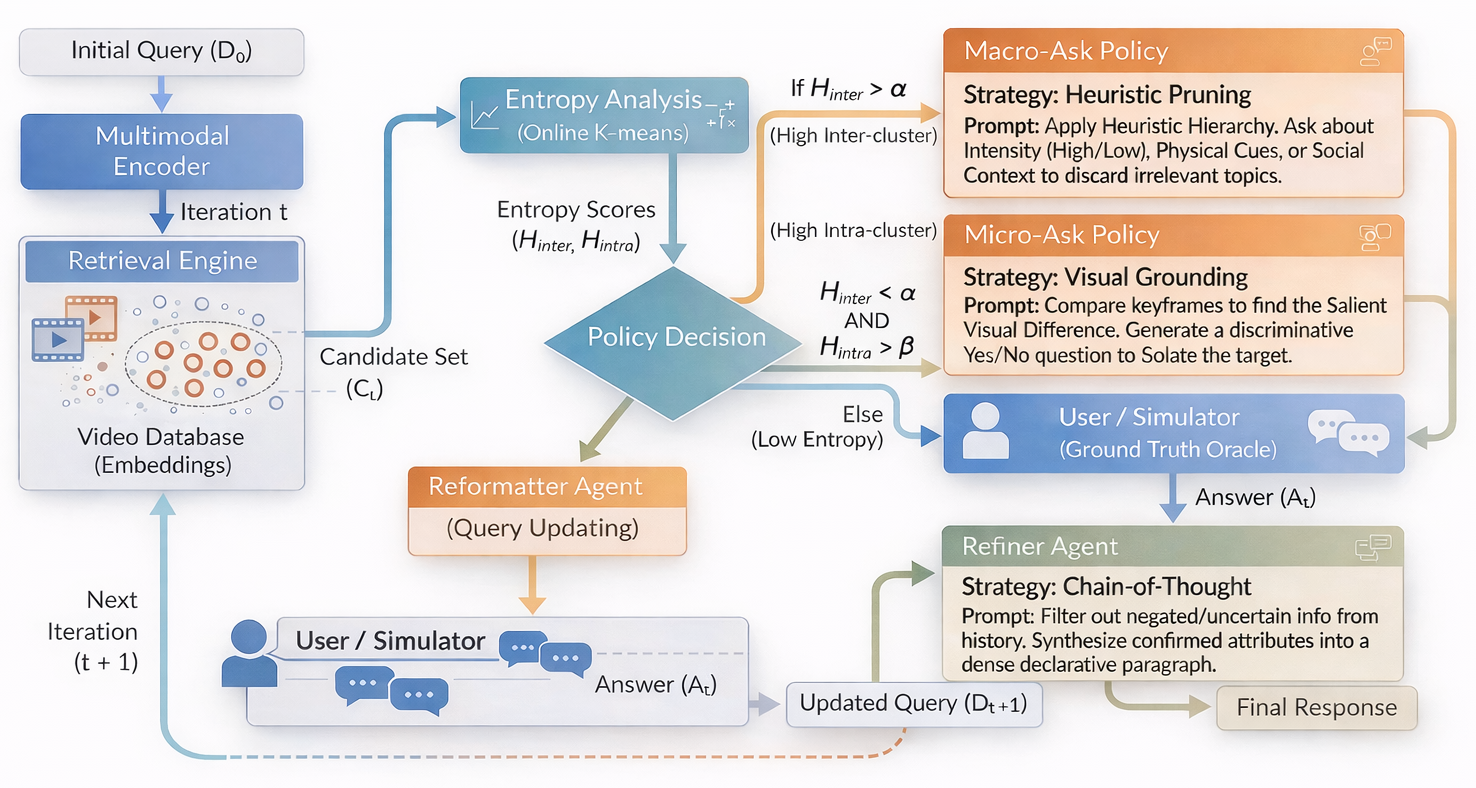}
  \caption{Overview of the ADEPT framework. The agent operates as an entropy-driven \textbf{closed loop}, utilizing uncertainty metrics ($H_{inter}, H_{intra}$) to dynamically select optimal strategies. Specific \textbf{system prompts} for each VLM agent are detailed in the call-out boxes.}
  \label{fig:adept_framework}
\end{figure*}

As illustrated in Fig.~\ref{fig:adept_framework}, ADEPT operates as an iterative closed loop. Instead of static heuristics, it diagnoses the uncertainty of $C_t$ via entropy metrics to dynamically arbitrate between \textit{Macro-Ask}, \textit{Micro-Ask}, and \textit{Refine} strategies. 

\subsection{Entropy-Driven Dynamic Strategy Selection}
Since standard entropy is infeasible for continuous embeddings, we employ cosine distance as a geometric proxy for \textit{Differential Entropy}~\cite{shannon1948mathematical}, linking uncertainty to spatial variance. At round $t$, candidates $E_{t-1}$ are clustered into $m$ subsets via $K$-means to compute:

\begin{itemize}
    \setlength{\itemsep}{0pt}
    \setlength{\topsep}{0pt}
    \setlength{\parsep}{0pt}
    \item \textbf{Inter-cluster entropy ($H_{\text{inter}}$)} captures global thematic divergence via the maximum pairwise centroid distance:
    \begin{equation}
        H_{\text{inter}} = \max_{i\neq j} \left(1 - \frac{\mathbf{c}_i \cdot \mathbf{c}_j}{\|\mathbf{c}_i\| \|\mathbf{c}_j\|}\right)
    \end{equation}
    High $H_{\text{inter}}$ implies diverse, conflicting topics, triggering \textit{Macro-Ask}.

    \item \textbf{Intra-cluster entropy ($H_{\text{intra}}$)} measures local semantic ambiguity via the average embedding-to-centroid distance:
    \begin{equation}
        H_{\text{intra}} = \frac{1}{|E_{t-1}|} \sum_{i=1}^{m} \sum_{\mathbf{e}_k \in S_i} \left(1 - \frac{\mathbf{e}_k \cdot \mathbf{c}_i}{\|\mathbf{e}_k\| \|\mathbf{c}_i\|}\right)
    \end{equation}
    High $H_{\text{intra}}$ indicates shared topics with detail confusion, triggering \textit{Micro-Ask}.
\end{itemize}

These metrics drive the policy decision (see Fig.~\ref{fig:adept_framework}) to dynamically select the optimal strategy.

\subsection{ASK Strategy}
When the decision engine selects ASK, the system poses questions to reduce uncertainty.

\subsubsection{Macro-Ask}
When $H_{\text{inter}} > \alpha$, the candidate set is thematically diverse, making fine-grained questioning inefficient. Macro-Ask addresses this by posing high-level questions to narrow the search space, using a heuristic hierarchy from abstract to concrete and focusing on \textit{intensity}, \textit{physical manifestation}, and \textit{social context} for disambiguation.

\subsubsection{Micro-Ask}
When $H_{\text{inter}} < \alpha$ but $H_{\text{intra}}>\beta$, retrieval has converged on a single topic yet candidates remain visually similar. Micro-Ask resolves this via two steps:
\begin{itemize}
    \setlength{\itemsep}{0pt}
    \setlength{\topsep}{0pt}
    \setlength{\parsep}{0pt}
    \item \textbf{Keyframe Selection:} sample frames at \textit{25\%, 50\%, 75\%} from each video and choose the one most similar to the video’s global embedding as keyframe.
    \item \textbf{Visual-Grounded Questioning:} process the keyframes in a VLM to detect objects, identify the most distinctive difference.
\end{itemize}

\subsection{Reformatter and Refiner}
\label{ssec:reformatter}
After acquiring new Q\&A information, query updates are managed by two specialized modules.

\subsubsection{Reformatter}
Triggered after each \textit{ASK} execution, this module takes the previous query $D_{t-1}$ and the current round’s $(Q_{t}, A_{t})$ pair as input. It seamlessly integrates the new attribute information into the existing query description to produce a richer and more precise intermediate query $D_{t}$.

\subsubsection{Refiner}
Triggered only when uncertainty converges ($H_{\text{inter}} \le \alpha, H_{\text{intra}} \le \beta$), this module takes the complete dialogue history $\mathcal{L}_{t-1}$ as input. Specifically, it employs a Chain-of-Thought (CoT) prompt to reason through the history, filtering out negated information (e.g., "not a blue shirt") before consolidating confirmed facts into a concise final query for the last retrieval step.

\section{Experimental Results}
\label{sec:typestyle}
This section presents a comprehensive empirical evaluation of the proposed ADEPT framework.

\subsection{Experimental Setup}
\subsubsection{Datasets and Evaluation Metrics}
To evaluate our model on tasks requiring fine-grained emotional and semantic understanding, we adopt two challenging test datasets.
\begin{itemize}
    \item \textbf{MAFW}~\cite{liu2022mafw}: Large-scale multimodal dataset for in-the-wild dynamic facial expression recognition (10,045 clips). We performed stratified sampling to create an 850-video evaluation subset.
    \item \textbf{MER2024}~\cite{lian2024mer}: Multimodal emotion recognition benchmark. We select a 332-video subset with human-annotated captions.
\end{itemize}

To encode video and text, we use the Google Multimodal Embedding API\footnote{\url{https://cloud.google.com/vertex-ai/generative-ai}}. All agent components (questioner, answerer, reformatter, and refiner) are implemented using Qwen2.5-VL~\cite{bai2025qwen2}. Retrieval performance is evaluated using the standard Recall@k metric.

\subsection{Baseline Models}
To evaluate our method, we selected four representative baseline models, spanning the spectrum from classic non-interactive methods to state-of-the-art general-purpose video understanding models.

\subsection{Hyperparameter Optimization}
To optimize hyperparameters $(m, \alpha, \beta)$, we employ a three-stage grid search on strictly partitioned data to prevent leakage. We first define a compact search space from the empirical entropy distributions of an analysis set, then conduct the final search on a separate validation set. The optimal settings are summarized in Table~\ref{tab:optimal_hyperparams}.

\begin{table}[t!]
\renewcommand{\arraystretch}{1.15}
\caption{Optimal hyperparameter configurations}
\label{tab:optimal_hyperparams}
\centering
\begin{tabular}{@{}lccc@{}} 
\toprule
\textbf{Dataset} & \textbf{$m$} & \textbf{$\alpha$} & \textbf{$\beta$} \\
\midrule
MAFW    & 8 & 0.0075 & 0.062 \\
MER2024 & 4 & 0.006  & 0.008 \\
\bottomrule
\end{tabular}
\end{table}

\subsection{Results}
As shown in Table \ref{tab:retrieval_results_split}, ADEPT consistently outperforms all baselines across datasets, highlighting the advantage of multi-turn interaction and the dynamic entropy-driven strategy.
\begin{table}[t!]
\centering

\small
\setlength{\tabcolsep}{2pt} 

\caption{Retrieval accuracy (\%); larger is better. 0 denotes single-shot retrieval without interaction rounds.}
\label{tab:retrieval_results_split}

\begin{tabularx}{\columnwidth}{@{}l c Y Y Y@{}}
\toprule
\multirow{2}{*}{\textbf{Model}} & \multirow{2}{*}{\textbf{Rounds}} & \multicolumn{3}{c}{\textbf{MAFW (Accuracy \%)}} \\ 
\cmidrule(l){3-5}
 & & R@1 & R@5 & R@10 \\
\midrule
Emotion-LLaMA (\cite{cheng2024emotion})   & 0 & 1.76 & 5.06  & 16.17 \\
CLIP4Clip (\cite{luo2021clip4clip})       & 0 & 6.59 & 15.65 & 21.65 \\
LLaVA-NeXT-Video (\cite{li2024llava})     & 5 & 12.70& 29.80 & 42.40 \\
BLIP + Heuristic (\cite{liang2023simple}) & 5 & 9.83 & 20.03 & 26.45 \\
\midrule
\multicolumn{5}{l}{\textit{Our Method}} \\
\hspace{1em} Round 0 (Single-shot) & 0 & 8.13  & 17.08 & 24.26 \\
\hspace{1em} Round 1               & 1 & 18.49 & 32.16 & 39.46 \\
\hspace{1em} Round 2               & 2 & 23.44 & 36.28 & 43.82 \\
\hspace{1em} Round 3               & 3 & 25.32 & 38.99 & 44.82 \\
\hspace{1em} Round 4               & 4 & 26.38 & 38.16 & 45.82 \\
\hspace{1em} Round 5               & 5 & \textbf{26.74} & \textbf{38.75} & \textbf{45.82} \\

\midrule
\addlinespace[2pt]
\midrule 
\multirow{2}{*}{\textbf{Model}} & \multirow{2}{*}{\textbf{Rounds}} & \multicolumn{3}{c}{\textbf{MER2024 (Accuracy \%)}} \\
\cmidrule(l){3-5}
 & & R@1 & R@5 & R@10 \\
\midrule
Emotion-LLaMA (\cite{cheng2024emotion})   & 0 & 0.91 & 7.27  & 17.64 \\
CLIP4Clip (\cite{luo2021clip4clip})       & 0 & 15.06& 35.24 & 43.37 \\
LLaVA-NeXT-Video (\cite{li2024llava})     & 5 & 26.81& 30.72 & 34.64 \\
BLIP + Heuristic (\cite{liang2023simple}) & 5 & 8.13 & 14.16 & 21.08 \\
\midrule
\multicolumn{5}{l}{\textit{Our Method}} \\
\hspace{1em} Round 0               & 0 & 16.57 & 35.24 & 44.88 \\
\hspace{1em} Round 1               & 1 & 18.67 & 28.01 & 35.24 \\
\hspace{1em} Round 2               & 2 & 22.89 & 33.13 & 42.17 \\
\hspace{1em} Round 3               & 3 & 23.00 & 34.04 & 43.37 \\
\hspace{1em} Round 4               & 4 & 24.40 & 36.14 & 44.88 \\
\hspace{1em} Round 5               & 5 & \textbf{24.40} & \textbf{35.84} & \textbf{46.99} \\
\bottomrule
\end{tabularx}
\end{table}

\subsection{Analysis}
\noindent\textbf{Adaptive Strategy \& Error Correction} 

\begin{figure}[t!]
  \centering
  \includegraphics[width=0.95\columnwidth]{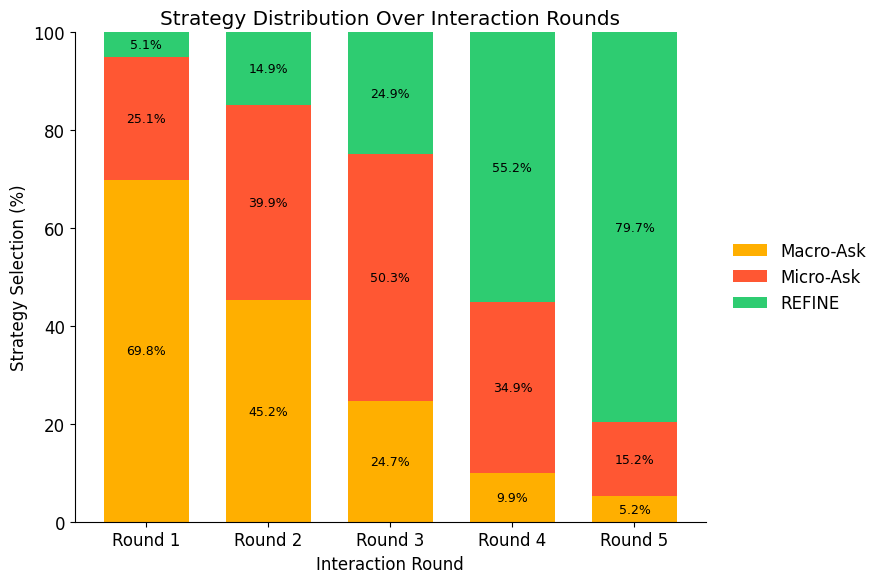}
  \vspace{-1mm}
  \includegraphics[width=0.95\columnwidth]{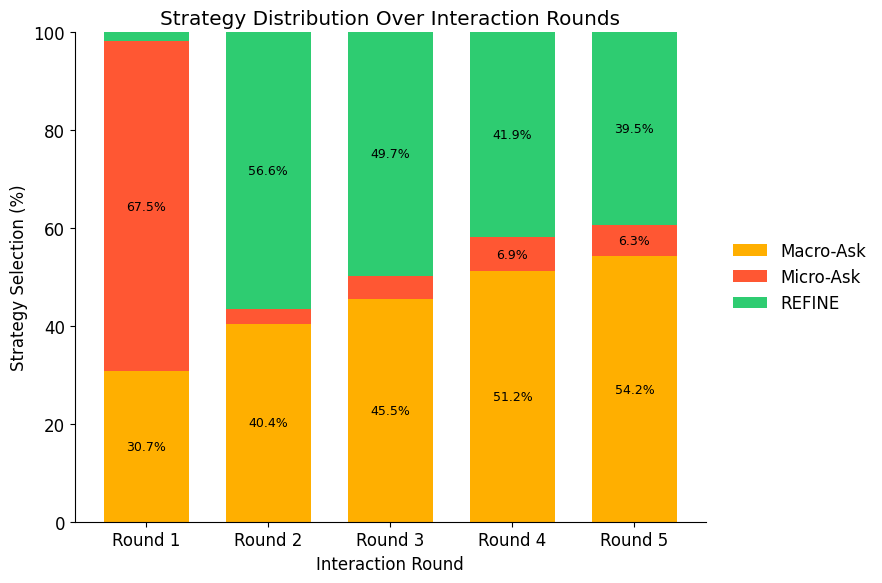}
  \caption{Strategy distributions for MAFW (top) and MER2024 (bottom).}
  \label{fig:strategy_distribution}
\end{figure}

Fig.~\ref{fig:strategy_distribution} reveals distinct behaviors: MAFW follows a linear \textit{Macro}$\to$\textit{Micro}$\to$\textit{Refine} path, while MER2024 triggers early \textit{Refine} (Round 2) to handle open-vocabulary noise. This adaptability yields massive gains for low-ranked videos ($>100$) on MAFW (Table~\ref{tab:rank_improvement_groups_combined}). Despite slight noise sensitivity in top-ranked MER2024 items, the framework effectively rescues mid-to-low ranked targets from retrieval failure.

\noindent\textbf{Parameter Sensitivity Analysis} 

Quantitative analysis highlights dataset-dependent sensitivity. Cluster count $m$ is decisive for discrete-label MAFW (importance score 0.53) to ensure valid partitioning. Conversely, $m$ is negligible (0.041) for open-vocabulary MER2024, where threshold $\alpha$ dominates. This confirms ADEPT prioritizes broad pruning in diverse spaces. Future work may further mitigate $m$-dependency by replacing K-means with soft clustering (e.g., GMMs) to better model overlapping semantics.

\begin{table}[t!]
\centering
\small
\setlength{\tabcolsep}{2.2pt}
\renewcommand{\arraystretch}{1.02}
\caption{Retrieval performance by initial-rank groups.
Init. = Initial-rank group; $N$ = Number of videos;
Rank$_\text{0}$ / Rank$_\text{5}$ = Average initial / final rank (after Round~5);
$\Delta$Rank = Rank$_\text{0}$ -- Rank$_\text{5}$;
Imp. (\%) = Improvement rate.}
\label{tab:rank_improvement_groups_combined}

\begin{tabularx}{\columnwidth}{@{}*{5}{>{\centering\arraybackslash}X} >{\centering\arraybackslash}p{1.55cm}@{}}
\toprule
\textbf{Init.} & \textbf{$N$} & \textbf{Rank$_0$} & \textbf{Rank$_5$} & \textbf{$\Delta$Rank} & \textbf{\mbox{Imp.\ (\%)}} \\
\midrule
\multicolumn{6}{l}{\textit{MAFW}}\\
1--10   & 206 & 5.82   & 1.21  & 4.61   & 79.21 \\
11--50  & 199 & 31.45  & 7.91  & 23.54  & 74.85 \\
51--100 & 107 & 73.19  & 14.52 & 58.67  & 80.16 \\
$>$100  & 337 & 240.51 & 35.84 & 204.67 & 85.10 \\
\midrule
\multicolumn{6}{l}{\textit{MER2024}}\\
1--10    & 156 & 3.58   & 5.12  & -1.54  & -43.02 \\
11--50   &  77 & 27.44  & 25.38 &  2.06  &  7.51 \\
51--100  &  41 & 71.46  & 55.49 & 15.98  & 22.35 \\
$>\!100$ &  58 & 174.48 & 90.66 & 83.83  & 48.04 \\
\bottomrule
\end{tabularx}
\end{table}

\section{Conclusion}
\label{sec:conclusion}
We present ADEPT, a training-free, entropy-driven agent designed to bridge the intent–-query gap. By diagnosing uncertainty via inter- and intra-cluster entropy, ADEPT dynamically switches strategies to efficiently guide interaction, establishing a principled baseline for interpretable active retrieval.

\vspace{0.3em}
\noindent\textbf{Limitations \& Future Work.}
ADEPT currently relies on offline parameter tuning and K-means clustering, which may limit generalization in overlapping semantic spaces. Future work will extend validation to more diverse benchmarks (e.g., MSR-VTT~\cite{xu2016msr}) and replace manual thresholds with a lightweight multi-armed bandit (MAB) policy~\cite{tang2025mba} for adaptive, online strategy selection.

\clearpage

\bibliographystyle{IEEEbib}
\bibliography{strings,refs}

@inproceedings{liu2022mafw,
  title={{MAFW}: A large-scale, multi-modal, compound affective database for dynamic facial expression recognition in the wild},
  author={Liu, Yuanyuan and Dai, Wei and Feng, Chuanxu and Wang, Wenbin and Yin, Guanghao and Zeng, Jiabei and Shan, Shiguang},
  booktitle={Proceedings of the 30th ACM International Conference on Multimedia},
  pages={24--32},
  year={2022}
}

@inproceedings{lian2024mer,
  title={{MER} 2024: Semi-supervised learning, noise robustness, and open-vocabulary multimodal emotion recognition},
  author={Lian, Zheng and Sun, Haiyang and Sun, Licai and Wen, Zhuofan and Zhang, Siyuan and Chen, Shun and Gu, Hao and Zhao, Jinming and Ma, Ziyang and Chen, Xie and others},
  booktitle={Proceedings of the 2nd International Workshop on Multimodal and Responsible Affective Computing},
  pages={41--48},
  year={2024}
}

@article{luo2021clip4clip,
  title={{CLIP4CLIP}: An Empirical Study of CLIP for End to End Video Clip Retrieval},
  author={Huaishao Luo and Lei Ji and Ming Zhong and Yang Chen and Wen Lei and Nan Duan and Tianrui Li},
  journal={Neurocomputing},
  year={2021},
  volume={508},
  pages={293-304}
}

@article{cheng2024emotion,
  title={{Emotion-LLaMA}: Multimodal emotion recognition and reasoning with instruction tuning},
  author={Cheng, Zebang and Cheng, Zhi-Qi and He, Jun-Yan and Wang, Kai and Lin, Yuxiang and Lian, Zheng and Peng, Xiaojiang and Hauptmann, Alexander},
  journal={Advances in Neural Information Processing Systems},
  volume={37},
  pages={110805--110853},
  year={2024}
}

@article{li2024llava,
  title={{LLaVA-NeXT-Interleave}: Tackling multi-image, video, and 3d in large multimodal models},
  author={Li, Feng and Zhang, Renrui and Zhang, Hao and Zhang, Yuanhan and Li, Bo and Li, Wei and Ma, Zejun and Li, Chunyuan},
  journal={arXiv preprint arXiv:2407.07895},
  year={2024}
}

@inproceedings{liang2023simple,
  title={Simple baselines for interactive video retrieval with questions and answers},
  author={Liang, Kaiqu and Albanie, Samuel},
  booktitle={Proceedings of the IEEE/CVF International Conference on Computer Vision},
  pages={11091--11101},
  year={2023}
}

@article{bai2025qwen2,
  title={{Qwen2.5-VL} Technical Report},
  author={Bai, Shuai and Chen, Keqin and Liu, Xuejing and Wang, Jialin and Ge, Wenbin and Song, Sibo and Dang, Kai and Wang, Peng and Wang, Shijie and Tang, Jun and others},
  journal={arXiv preprint arXiv:2502.13923},
  year={2025}
}

@inproceedings{ma2022x,
  title={{X-CLIP}: End-to-end multi-grained contrastive learning for video-text retrieval},
  author={Ma, Yiwei and Xu, Guohai and Sun, Xiaoshuai and Yan, Ming and Zhang, Ji and Ji, Rongrong},
  booktitle={Proceedings of the 30th ACM International Conference on Multimedia},
  pages={638--647},
  year={2022}
}

@inproceedings{maeoki2020interactive,
  title={Interactive video retrieval with dialog},
  author={Maeoki, Sho and Uehara, Kohei and Harada, Tatsuya},
  booktitle={Proceedings of the IEEE/CVF Conference on Computer Vision and Pattern Recognition Workshops},
  pages={952--953},
  year={2020}
}

@inproceedings{han2024merlin,
    title = "{MERLIN}: Multimodal Embedding Refinement via {LLM}-based Iterative Navigation for Text-Video Retrieval-Rerank Pipeline",
    author = "Han, Donghoon  and
      Park, Eunhwan  and
      Lee, Gisang  and
      Lee, Adam  and
      Kwak, Nojun",
    editor = "Dernoncourt, Franck  and
      Preo{\c{t}}iuc-Pietro, Daniel  and
      Shimorina, Anastasia",
    booktitle = "Proceedings of the 2024 Conference on Empirical Methods in Natural Language Processing: Industry Track",
    month = nov,
    year = "2024",
    address = "Miami, Florida, US",
    publisher = "Association for Computational Linguistics",
    url = "https://aclanthology.org/2024.emnlp-industry.41/",
    doi = "10.18653/v1/2024.emnlp-industry.41",
    pages = "547--562"
}

@inproceedings{fan2024videoagent,
  title={Videoagent: A memory-augmented multimodal agent for video understanding},
  author={Fan, Yue and Ma, Xiaojian and Wu, Rujie and Du, Yuntao and Li, Jiaqi and Gao, Zhi and Li, Qing},
  booktitle={Proceedings of the European Conference on Computer Vision},
  pages={75--92},
  year={2024},
}

@inproceedings{shang2024traveler,
    title = "{T}rave{LER}: A Modular Multi-{LMM} Agent Framework for Video Question-Answering",
    author = "Shang, Chuyi  and
      You, Amos  and
      Subramanian, Sanjay  and
      Darrell, Trevor  and
      Herzig, Roei",
    editor = "Al-Onaizan, Yaser  and
      Bansal, Mohit  and
      Chen, Yun-Nung",
    booktitle = "Proceedings of the 2024 Conference on Empirical Methods in Natural Language Processing",
    month = nov,
    year = "2024",
    address = "Miami, Florida, USA",
    publisher = "Association for Computational Linguistics",
    url = "https://aclanthology.org/2024.emnlp-main.544/",
    doi = "10.18653/v1/2024.emnlp-main.544",
    pages = "9740--9766"
}

@inproceedings{thawakar2024composed,
  title={Composed video retrieval via enriched context and discriminative embeddings},
  author={Thawakar, Omkar and Naseer, Muzammal and Anwer, Rao Muhammad and Khan, Salman and Felsberg, Michael and Shah, Mubarak and Khan, Fahad Shahbaz},
  booktitle={Proceedings of the IEEE/CVF Conference on Computer Vision and Pattern Recognition},
  pages={26896--26906},
  year={2024}
}

@article{shannon1948mathematical,
  title={A mathematical theory of communication},
  author={Shannon, Claude E},
  journal={The Bell system technical journal},
  volume={27},
  number={3},
  pages={379--423},
  year={1948},
  publisher={Nokia Bell Labs}
}

@inproceedings{xu2016msr,
  title={MSR-VTT: A large video description dataset for bridging video and language},
  author={Xu, Jun and Mei, Tao and Yao, Ting and Rui, Yong},
  booktitle={Proceedings of the IEEE Conference on Computer Vision and Pattern Recognition},
  pages={5288--5296},
  year={2016}
}

@inproceedings{tang2025mba,
  title={{MBA-RAG}: a bandit approach for adaptive retrieval-augmented generation through question complexity},
  author={Tang, Xiaqiang and Gao, Qiang and Li, Jian and Du, Nan and Li, Qi and Xie, Sihong},
  booktitle={Proceedings of the 31st International Conference on Computational Linguistics},
  pages={3248--3254},
  year={2025}
}

\section*{Compliance with Ethical Standards}
This research did not involve human participants or animals and therefore does not require ethical approval. All datasets used in this study are publicly available.

\end{document}